# Considering Avatar Crossing as Harm or Help for Adolescents in Social VR

JAKKI O. BAILEY, School of Information, University of Texas at Austin, USA

XINYUE (SALLY) YOU, School of Information, University of Texas at Austin, USA

## 1 INTRODUCTION

People leverage avatars to communicate nonverbal behaviors in immersive virtual reality (VR), like interpersonal distance [2, 6] and virtual touch [5]. However, violations of appropriate physical distancing and unsolicited intimate touching behavior in social virtual worlds represent potential social and psychological virtual harm to older adolescent users [4, 8]. Obtaining peer acceptance and social rewards, while avoiding social rejection can drive older adolescent behavior even in simulated virtual spaces [1, 3], and while "the beginning of adolescence is largely defined by a biological event, [...] the end of adolescence is often defined socially" [3] (p.912). Avatar crossing, the phenomenon of avatars walking through each other in virtual environments, is a unique capability of virtual embodiment, and offers intriguing possibilities and ethical concerns for older adolescents experiencing social virtual spaces. For example, the ability to cross through and share positions with other avatars in a virtual classroom helps students concentrate on accessing and comprehending information without concerns about blocking others when navigating for better viewpoints [10]. However, the ability to cross through others in virtual spaces has been associated with a reduction in perceived presence and avatar realism, coupled with a greater level of discomfort and intimidation in comparison to avatar collisions [12]. In this article, we consider the potential benefits and harms of utilizing avatar crossing with adolescent users.

## 2 AVATAR CROSSING FOR REDUCING HARM AMONG ADOLESCENTS IN VIRTUAL WORLDS

Adolescence is a time of rapid brain development for processing social information, and social stress, like ostracism, experienced during adolescence may be longer lasting and qualitatively different form of stress than other periods of life [3]. According to Blakemore and Mills [3], the behaviors characteristic of adolescence "such as heightened self-consciousness, mood variability, novelty seeking, risk taking, and peer orientation, are fundamental to the successful transition into a stable adult role" (p. 9.11). These behaviors, while important for the shift to adulthood, may increase adolescents' likelihood of engaging in or experiencing harmful activities in embodied and immersive social virtual worlds. For example, older children and adolescents report harassment, bullying, and stalking behaviors in online social virtual worlds [7, 8]. In addition, adults and children co-mingle in these social embodied and immersive virtual spaces, creating additionally complex social dynamics, such as exposure to mature content [7].

Within VR settings, individuals often map their body schema onto their avatars (i.e., virtual embodiment) and feel ownership over those bodies [9, 11]. Furthermore, the act of virtual embodiment can influence users' behaviors and interactions, leading to shifts in self-perception both within and outside of VR spaces [9]. With avatars experienced as the self, the potential harm to adolescents in embodied and immersive social virtual worlds need to be specifically considered. The negative impact of unwanted touching in is a major concern for VR embodiment in social virtual worlds. For instance, adult users experienced physical distress when their avatar was slapped by another person [11]. Avatar crossing may act as a solution to mitigate the negative impact of unsolicited and unwanted virtual touching for young users. By removing physics from avatar bodies and allowing avatar crossing, adolescent users can avoid the sensation of experiencing direct contact to their virtual bodies. However, in extreme case of avatar crossing, users can enter the virtual body of another, which may adversely affect younger users. Without safeguards that allow adolescent users to control when avatar crossing happens, they may feel a different yet still powerful level of discomfort from online social interactions.

The adolescent perception of avatar crossing and overlapping remains uncertain, indicating the need for further investigation into their nuanced implications. Ultimately, continued research in this space can lead to policies and technologies that better capture how we understand personal space and acceptable proxemic behaviors in embodied and immersive virtual spaces, while leveraging the unique capabilities of VR to enhance young users' social experiences.